\documentclass{aa}  
\usepackage{natbib}
\bibpunct{(}{)}{;}{a}{}{,} 

\usepackage{graphicx}
\usepackage{longtable}
\usepackage{supertabular}
\usepackage{subcaption}
\usepackage{xspace}
\newcommand{\fluence}{\,Jy\,ms\xspace} % cgs units for DM
\newcommand{\dm}{\,pc\,cm$^{-3}$\xspace} % cgs units for DM
\newcommand{\xte}{XTE\,J1810$-$197\xspace} % source

\newcommand{\bause}{2024A&A...686A.144B}

\usepackage{txfonts}
\begin{document} 
%% The lineno packages adds line numbers. Start line numbering with
%% \begin{linenumbers}, end it with \end{linenumbers}. Or switch it on
%% for the whole article with \linenumbers after \end{frontmatter}.

%\begin{linenumbers}
%\nolinenumbers

%\switchlinenumbers

   \title{Giant radio pulses in the magnetar XTE J1810$-$197 detected with the IAR's telescopes}

   \author{S. B. Araujo Furlan\inst{1,2} \fnmsep\thanks{E-mail: susana.araujo@unc.edu.ar}
          \and
          G. E. Romero\inst{3,4}
          \and
          E. Zubieta\inst{3,4}
           \and
          G. Gancio\inst{3}
          \and 
          F. Garc\'{i}a\inst{3,4}
          \and 
          S. del Palacio \inst{5}
          \and
          C. O. Lousto \inst{6}
          }

   \institute{ Instituto de Astronom\'\i{}a Teórica y Experimental, CONICET-UNC, Laprida 854, X5000BGR – Córdoba, Argentina
         \and
            Facultad de Matemática, Astronomía, Física y Computación, UNC. Av. Medina Allende s/n , Ciudad Universitaria, CP:X5000HUA - Córdoba, Argentina.
        \and 
        Instituto Argentino de Radioastronom\'ia (CCT La Plata, CONICET; CICPBA; UNLP),
              C.C.5, (1894) Villa Elisa, Buenos Aires, Argentina
        \and
               Facultad de Ciencias Astron\'omicas y Geof\'{\i}sicas, Universidad Nacional de La Plata, Paseo del Bosque, B1900FWA La Plata, Argentina
               \and
               Department of Space, Earth and Environment, Chalmers University of Technology, SE-412 96 Gothenburg, Sweden
               \and
               Center for Computational Relativity and Gravitation, Rochester Institute of Technology, 85 Lomb Memorial Drive, Rochester, New York 14623, USA
             }

   \date{Received  23 July 2025; accepted 26 February 2026}

% \abstract{}{}{}{}{} 
% 5 {} token are mandatory
 
  \abstract
   {\xte is one of the six magnetars that exhibit radio emission, sometimes visible only during periods of increased activity or outbursts. \xte underwent its latest outburst in December 2018.}
   {We aim to analyze the population of single pulses of \xte in radio, characterize its timing properties, and explore the potential connection between magnetars and fast radio bursts (FRBs).}
    {We observed \xte between 29 September 2022 and 14 July 2023 with the radio telescopes at the Argentine Institute of Radioastronomy (IAR). We searched for single pulses in time series at a DM range of 100--400\dm, with a threshold in signal-to-noise ratio (S/N) of 8. We folded each observation to obtain an integrated pulse profile. We also analyzed archival X-ray observations of the MAXI instrument from the same period, and studied the flux evolution and the magnetar's activity.}
   {We found 249 giant pulses at a DM mean value of $178.8\pm0.1$\dm. We measured peak flux densities up to 119~Jy, and fluences up to 58\fluence. We fitted a power law distribution to the flux density, obtaining an index of $-4.0\pm0.3$. We observed a maximum rate of approximately 15 pulses per hour on 20 February 2023, followed by an abrupt disappearance of transient radio emission, indicating a transition to a less active state. 
   The brightest single pulses are limited to a $\sim$2\% of the rotational phase and have similar fluence values to the reported intermediate FRB-like bursts of SGR\,1935+2154.
   No significant X-ray activity in the MAXI data was detected during the radio observing period. }
   {This is the first study of single radio pulses of a magnetar using IAR data, showing the potential of the upgraded telescopes for investigating the transient radio sky. The properties of the single pulses detected here show the magnetar transient nature and capability to emit high-luminosity pulses. We compared the detected emission to FRB-like bursts and single pulses emitted by SGR\,1935+2154. Even though the mechanism producing all the events should be coherent, the luminosity of the events, features on the dynamic spectra, and the difference between being phase confined or not, indicate that \xte presents GP emission, while SGR\,1935+2154 only shows normal single pulses or FRB bursts. This could indicate that the conditions for producing each type of event differ.}

   \keywords{stars: magnetars--
                stars: neutron --
                stars: individual: \xte --
                radio continuum: stars--
                methods: observational --
                methods: data analysis
               }

   \maketitle

\section{Introduction}

Magnetars are neutron stars with extremely strong surface magnetic fields ($B\sim10^{13}$--$10^{15}\,\mathrm{G}$). They are among the youngest and slowest neutron stars, with periods of $\sim$1 to 12~s \citep{2014ApJS..212....6O}. They display a rich transient phenomenology, including giant flares, short bursts, and outbursts, detected mainly at X-rays and $\gamma$-rays \citep{kaspi_magnetars_2017}. The energy for the emission is provided by the decaying magnetic fields, which could be composed of a twisted dipole plus some local twisted multipole field \citep{duncan_formation_1992, tong_magnetospheric_2023}. 

Only 6 out of the 30 known magnetars have been detected in radio so far \citep{2014ApJS..212....6O}\footnote{\url{https://www.physics.mcgill.ca/~pulsar/magnetar/main.html}}. 
Five of them presented transient radio pulsations, in general associated with X-ray outbursts. The sixth one, SGR\,1935+2154, has been found to emit a fast radio burst (FRB) \citep{bochenek_fast_2020}  and FRB-like radio bursts. During its October 2022 outburst, bright radio bursts were detected within the 9-hour interval between the two glitches of an unusual double-glitch event \citep{maan_gbt_2022, dong_chimefrb_2022, hu_rapid_2024}. This radio activity may arise from the emergence of a short-lived magnetospheric wind, as proposed by \cite{hu_rapid_2024}. The detection of FRB 200428 from SGR\,1935+2154 strongly supports the hypothesis that magnetars can power at least some FRBs. These results turned the community's attention towards the conditions required for a magnetar to produce a FRB-like burst.

Studying the pulsed radio emission of magnetars provides an avenue to explore their unique spectral-temporal properties and phenomenology. This radio emission has been detected across a wide range of frequencies, observed as an integrated pulse profile and as bright single pulses. Furthermore, the radio emission exhibits transient episodes of activity, characterized by short-duration (e.g., daily) on-off switching that can occur even in the absence of concurrent X-ray activity, as exemplified by magnetar PSR J1622-4950 \citep{levin_radio-loud_2010}. It also shows high pulse-to-pulse variability both in single and in folded pulses (\citealp[for recent reviews]{kaspi_magnetars_2017,esposito_magnetars_2021}). 

\xte was the first magnetar detected at radio frequencies \citep{halpern_discovery_2005, camilo_transient_2006}. Here we present the results of a timing and single pulse analysis on the radio observations of \xte conducted daily with the radiotelescopes at the Argentine Institute of Radioastronomy (IAR) between September 2022 and July 2023. We also present an analysis of the X-ray activity during this period using archival observations of the \textit{Monitor of All-sky X-ray Image} (\textit{MAXI}). This work is the first single-pulse study of magnetar observations ever done with the IAR telescopes and the first studying the radio activity of the magnetar in the mentioned epoch. 

The structure of the paper is as follows. In Sect.~2 we review previous observational results associated with this magnetar. In Sect.~3 we describe the observations obtained in this work. In Sect.~4 we present the reduction and data analysis. In Sects.~5 and 6, we show the results obtained from the radio and X-ray observations, which we discuss in Sect.~7. We summarize our main conclusions in Sect.~8.

%--------------------------------------------------------------------
\section{The magnetar \xte}

\xte was discovered serendipitously in X-rays on 15 July 2003 by the \textit{Rossi X-ray Timing Explorer} (\textit{RXTE}) while observing a soft gamma repeater. It had a transient X-ray outburst between 17 November 2002--23 January 2003. With the study of its spin-down rate and slow period ($\sim$5.54~s), it showed the first signatures of being a magnetar \citep{ibrahim_discovery_2004}. It was the first of its class with detected radio emission, initially as a point source in 2004 interferometric observations \citep{halpern_discovery_2005} and later via radio pulses in single-dish observations beginning in 2006, revealing its transient nature \citep{camilo_transient_2006}. This is known in the literature as the first radio and X-ray outburst of the source.

\cite{camilo_transient_2006} showed that the magnetar emitted bright, narrow and highly linearly polarized pulses in each rotation, suggesting that a magnetar could have pulsar-like emission. They noted that the observed radio emission requires a coherent mechanism, which is likely powered by pair production enabled by the strong voltage induced in the magnetosphere. While the resulting radio beam may be narrow and miss our line of sight, its activation is linked to magnetospheric twisting. As reviewed by \cite{kaspi_magnetars_2017}, the emission is thought to originate from long-lived currents produced by the magnetosphere untwisting, along magnetic field lines in the magnetosphere far from the star. In the 2006 study, they detected single pulses of width $\leq 10$~ms and flux densities of $\leq 10$~Jy. The period of detected radio activity went from 2004 until the end of 2008 \citep{camilo_radio_2016}. This last study showed the extremely unpredictable behavior of the radio emission of the magnetar. The pulse profile showed day-to-day variations, changing the number of components, and the radio spectrum changed from being flat in 2004 to a steeper one by 2008. In addition, the radio emission suddenly disappeared until the 2018 outburst. 

A second radio and X-ray outburst was detected in late 2018, with an onset between 26 October and 8 December 2018 in radio and between 20--26 November 2018 in X-rays \citep{2018ATel12284....1L,levin_spin_2019, gotthelf_2018_2019}. On the following days ---14, 19, and 20 December--- integrated radio pulse profiles were detected with the IAR's telescopes \citep{2018ATel12323....1D}. 
Several different radio studies were conducted following this
outburst. The evolution of the pulse profile and the behavior of the broadband spectra and polarisation were extensively studied \citep{2019ApJ...874L..14D,levin_spin_2019, eie_multi-frequency_2021, desvignes_freely_2024, lower_linear_2024}, including observations at frequencies as low as 300~MHz \citep{maan_magnetar_2022} and as high as 353~GHz \citep{torne_submillimeter_2022}. The main results obtained were that the pulse profile showed a daily changing behavior and a steeper spectrum \citep{torne_submillimeter_2022}. 

 Other studies focused on single pulse analysis \citep{maan_distinct_2019, 2022MNRAS.510.1996C, 2024A&A...686A.144B}. The latter two made an analysis on long-term observations and were able to analyze the change in the single pulse population throughout the period from 8 December 2018 until 14 November 2021, being \cite{2024A&A...686A.144B} the one with most recent observations.  An interesting result was the detection of bursts of emission in the form of giant pulses (GPs) or giant-pulse like \citep{maan_distinct_2019, 2024A&A...686A.144B, 2022MNRAS.510.1996C} from this magnetar. Clearly, there is no universal consensus on the definition of GPs. One thing on which there is agreement is that they are extremely energetic and occur in a narrow phase window compared to normal pulses. One working definition in previous studies on \xte is that the pulse's energy is 10 times the average energy. \cite{2022MNRAS.510.1996C} obtained the average pulse energy from the integrated profiles of only the rotations that did not present GP-like emission. We considered that all the detected single pulses are GPs (see Sect. \ref{sec:onlyGP}). To asses the energetics of the pulses, we introduce the fluence defined as $ S_{\mathrm{peak}} \times \tau$, where $S_{\mathrm{peak}}$ is the flux density of the peak and $\tau$ is the pulse duration\footnote{This quantity is also referred as the energy of the pulse in some works on GPs \citep{kuzmin_giant_2006,karuppusamy_giant_2010}.}.
 
 The observations with the shortest sampling time were taken with the Effelsberg telescope, with a sampling of $131 \mathrm{\,\mu s}$ \citep{2022MNRAS.510.1996C}, followed by the Stockert telescope with $218.45\mathrm{\,\mu s}$ \citep{2024A&A...686A.144B} and finally the GMRT with $655\,\mu$s \citep{maan_distinct_2019}. This is relevant when identifying the substructure and width of the pulses. 

We note that different works use slightly different definitions of a single pulse and its width depending on the method employed.  \cite{2022MNRAS.510.1996C} defines a single pulse as the emission detected within one rotation of the neutron star, presenting sometimes substructure or spiky emission. While \cite{2024A&A...686A.144B} defines a single pulse as the spiky emission, where they could have several single pulses in one rotation. \cite{2024A&A...686A.144B} found pulse widths in the range of 0.65--30~ms, with around 50\% of the pulses having a width of $\leq4\mathrm{\, ms}$. The flux densities they found ranged from just a few Jy to up to 300~Jy, with some outliers of 600~Jy.

Interestingly, the average fluxes in radio and X-rays are not correlated, as shown by \cite{2022MNRAS.510.1996C} during May--September 2020 and by \cite{2024A&A...686A.144B} during February 2021. This suggests that the activity in the radio regime may evolve independently of the X-ray activity.

\section{Observations}

\subsection{Radio observations}

We performed high-cadence observations of \xte with the two 30-m single dish antennas at IAR, "Carlos M. Varsavsky" (A1) and "Esteban Bajaja" (A2), located in Parque Pereyra Iraola, Argentina \citep{guille2020}. The monitoring started on 29 September 2022 (MJD 59851) and continued until 14 July 2023 (MJD 60139). Despite being close to the ecliptic in December, we note that the source was never behind the Moon nor the Sun.

Each antenna's receiver consists of two ETTUS board. For A1, the central frequency was 1424~MHz with a bandwidth of 112~MHz divided into 128 channels, on one circular polarization. In the case of A2, the central observing frequency was set to 1428~MHz with a bandwidth of 56~MHz divided into 64 channels, on both circular polarizations. The sample time in both cases was $146\,\mu \mathrm{s}$.
The data was recorded into filterbank files. Here we analyzed 149~days of observations obtained with A2. The A1 observations were excluded from the analysis due to a higher level of radio-frequency interferences (RFIs). The A2 observing times ranged between 18~min and 172~min, yielding a total of 269~h on source. We highlight that this high-cadence campaign covered approximately 52\% of the observing period.

\begin{figure}
    \centering
    \includegraphics[width=0.9\linewidth]{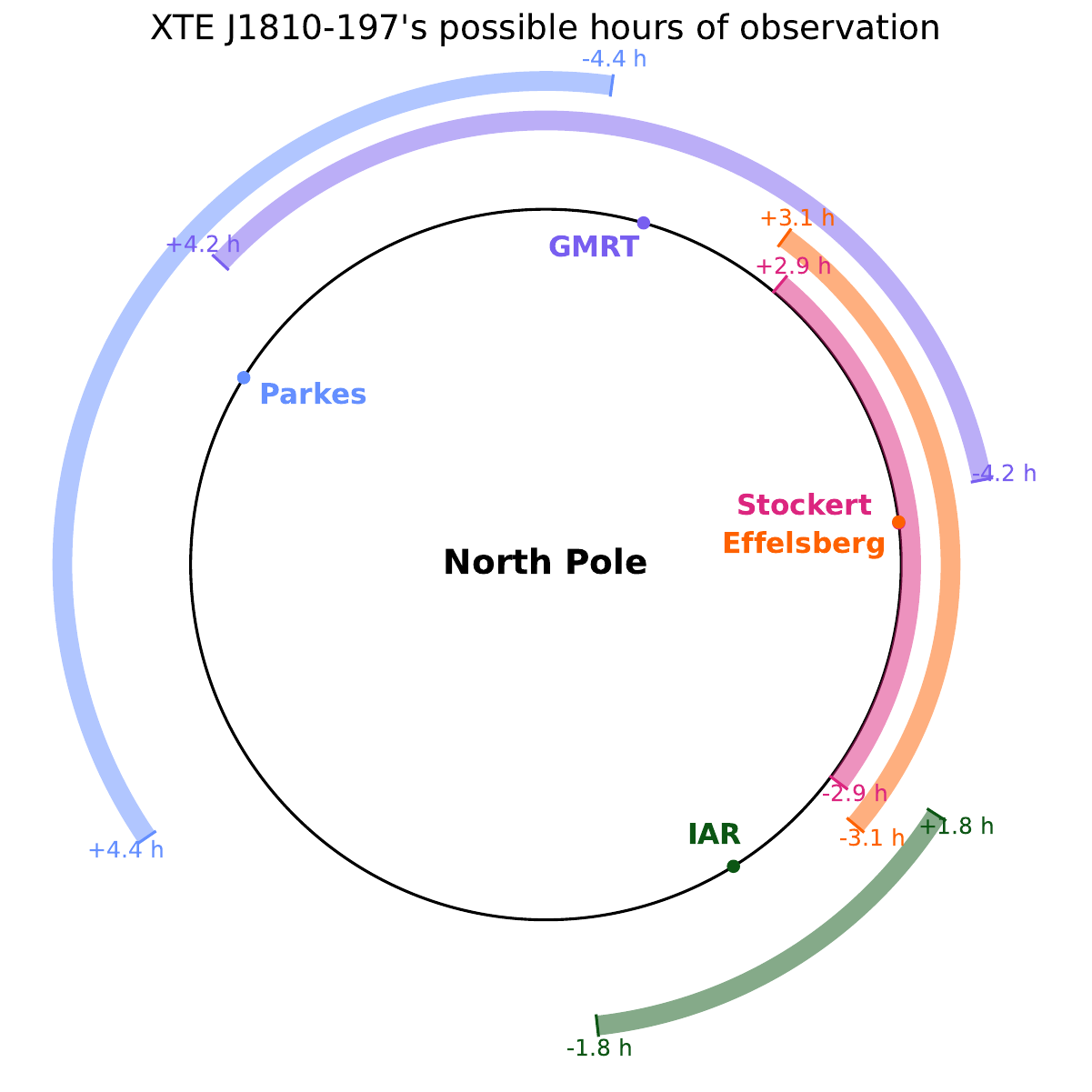}
    \caption{Observable time of \xte for different telescopes that have conducted high-cadence campaigns. The arc matching the telescope's color represents the source's possible observation time for that telescope. We highlight the very short overlap between the IAR and other observatories.}
    \label{fig:visibilidad}
\end{figure}

The geographic location of the IAR provides unique access to temporal windows unavailable to other radio telescopes during previous high-cadence campaigns. Figure\,\ref{fig:visibilidad} compares the possible hours on source for several telescopes (Parkes, GMRT, Stockert, Effelsberg), considering their elevation limit. As it is a southern source, IAR can observe \xte for almost 13 hours above the horizon, although mechanical limitations reduce its maximum tracking time to 3.6\,h. Figure\,\ref{fig:visibilidad} shows that high-cadence campaigns with several telescopes enable detection of intraday transients events, highlighting the benefit of geographically dispersed observations.

\subsubsection{Calibration}

In mid-2022, the IAR underwent an upgrade with the installation of a parallel digitalizer board consisting of Reconfigurable Open Architecture Computing Hardware (ROACH) boards. Virtually, the IAR antennas have two receivers that differ only on the digitalizer board, either ETTUS or ROACH. In August 2022, we started the calibration of the system employing the ROACH board, calibrating the noise diode MC7014 ENR 35dB \citep{calibDiodoDeRuido}. Following efforts on the calibration of the dual-polarization of A2 can be found in \cite{guilleRevistaMexicana}. We only have a noise temperature value for the ROACH board. 

The flux density values obtained in this work rely on two main assumptions: (1) the noise temperature was identical for both the ETTUS and the ROACH board, and (2) the noise temperature remained constant throughout the entire observation period. These assumptions are justified below.

Equation \ref{eq:gain_tnoise} describes the noise temperature  for a receiver

\begin{equation}\label{eq:gain_tnoise}
    T_{N} = T_{N1} +  \frac{T_{N2}}{G_{1}}+\frac{T_{N3}}{G_{1}G_{2}}+ ...+ \frac{T_{Nf}}{\prod^{(f-1)}_{i=1}G_{i}},
\end{equation}
where $T_{N}$ is the total noise temperature for a receiver composed of $f$ elements. $T_{N1}$  and $G_{1}$ are the individual noise temperature and gain of the first element, respectively, $T_{N2}$ and $G_{2}$ of the second element, and so on. 
 The contribution of the final element is minimal, as it is divided by the product of the gains of all preceding stages.  Consequently, the first element in the receiver chain dominates the overall noise temperature.
In our case, the final element was the digitalizer board, either the ROACH or the ETTUS. Both boards received the same input signal. Their noise temperature contribution was insignificant as the factor that multiplies them is of the order of $10^{-6}$.   

Since we only have systematic measurements of the noise temperature from mid-2023, covering nearly a year of daily data (from June 2023 to May 2024), we adopt the average value measured during this period: $T_{N} = 50 \pm 8\,\mathrm{K}$. The receiver system underwent no modifications during this time, supporting the assumption of stability. We therefore consider this value representative of the system from September 2022 to May 2024, with any potential variation reflected in the quoted uncertainty.

Thus, we obtained the flux density of a temporally resolved single pulse using the modified radiometer equation (e.g., \cite{maan_deep_2014}):
\begin{equation}
    S_{\mathrm{peak}} = (S/N)_{\mathrm{peak}} \frac{2k_{B}T_{\mathrm{sys}}}{A_{e}(z)\sqrt{n_{p}W \ \Delta\nu}},
\end{equation}
where  $T_{sys}$ is the system temperature in K, $A_{e}(z)$ is the effective collecting area as a function of the zenith angle (0.3 times the area of the 30-m dish for A2 in m$^2$), $\Delta \nu$ is the bandwidth in Hz ($56\times10^{6}$~Hz), $n_{p}$ is the number of polarizations (2 for A2), $(S/N)_{\mathrm{peak}}$ is the peak signal-to-noise ratio of the pulse, and $W$ the observed pulse width. In this work, we considered $T_{sys}$ as the noise temperature of the system. After cleaning for RFIs, we determined that one or two channels were fully masked in all observations and used $\Delta \nu$ as 0.9 the observed bandwidth ($0.9\times56\times10^6$~Hz).

\subsection{X-ray observations}

We analyzed archival X-ray observations of \xte to search for possible correlations with the radio emission. 
We studied the daily light curves of \xte provided by the \textit{MAXI} instrument \citep{matsuoka_maxi_2009}. \textit{MAXI} is a high-energy astrophysical experiment onboard the International Space Station (ISS). It comprises two semi-circular arc-shaped X-ray slit cameras with wide Field of View (FoV). It harbors two kinds of X-ray detectors that collect photons from the slit cameras: a gas proportional counter, the Gas Slit Camera (GSC; working in the 2--30~keV band), and an X-ray CCD, Solid-state Slit Camera (SSC; operating in the 0.5--12~keV range). Being on the ISS, it maps the entire sky on each orbit, every $\sim$92~min.

\textit{MAXI} provides their data already processed with different time binning. We used the 1-orbit light curves, and associated $\sim$15$^{\circ}$ FoV images, together with the all-sky maps to check possible source contamination\footnote{\url{http://maxi.riken.jp/v7l3h/J1809-197/}}. The observations retrieved spanned the period from 1 September 2022 (MJD 59823) to 28 April 2024 (MJD 60428).

\section{Reduction and data analysis}

\subsection{Radio}

The radio analysis involves two main parts. The first part was carried out by PuMA pipeline\footnote{\url{https://github.com/PuMA-Coll/PuMA}}, which employs PRESTO \citep{presto} as the core tool of its process, performing RFI masking and phase folding of the observations. This is described in detail in Section~\ref{section:ReductionTiming}. The second part, focused on transient activity, was performed with a bash pipeline developed for this work, described in Section~\ref{section:ReduccionSingle}. After processing all observations with these pipelines, we conducted an in-depth analysis with a Python-based script.

\subsubsection{Timing}\label{section:ReductionTiming}

We used the \textsc{rfifind} and \textsc{prepfold} tasks in PRESTO package to remove RFIs and fold the entire duration of each observation using a previously reported ephemeris to obtain integrated pulse profiles. Next, we used the task \textsc{pat} in PSRCHIVE \citep{hotan_psrchive_2004,van_straten_pulsar_2012} to calculate the Time of Arrival (TOA) using the Fourier phase gradient-matching method for template fitting. We used the pulse profile of an observation with high S/N to construct the template with the \textsc{psrsmooth} package in PSRCHIVE.

We developed a timing model to characterize the rotation of the magnetar by tracking the TOAs of its pulses. We used a Taylor expansion to describe the temporal evolution of the pulsar phase, given by Eq. (\ref{eq:timing-model}), 

\begin{equation}\label{eq:timing-model}
    \varphi(t)=\varphi+\nu(t-t_0)+\frac{1}{2}\dot{\nu}(t-t_0)^2+\frac{1}{6}\Ddot{\nu}(t-t_0)^3+...,
\end{equation}
where $\nu$, $\dot{\nu}$, and $\ddot{\nu}$ represent the rotation frequency of the magnetar and its first and second derivatives at the reference epoch $t_0$, and $\varphi$ is the phase of the pulsar at $t_0$. Various physical phenomena, such as glitches or sudden changes in the frequency derivative can be investigated through this method \citep{2022RPPh...85l6901A}.

We first obtained the parameters for the timing model from \cite{2016ApJ...820..110C}, and subsequently updated them by fitting our data to the timing model (Eq. \ref{eq:timing-model}) using the \textsc{Tempo2} software package \citep{2006MNRAS.369..655H}. 

To address jitter and systematic errors, we incorporated the parameters EFAC and EQUAD. These parameters model white noise by adjusting each TOA uncertainty as follows:

\begin{equation}
    \sigma_{\rm TOA} = \sqrt{{\rm EQUAD}^2 +( {\rm EFAC} \times \sigma_i)^2},
\end{equation}
where $\sigma_i$ represents the TOA uncertainty for each observation derived from the cross-correlation between the template profile and the folded observation. EFAC accounts for the unmodeled instrumental effects and inaccuracies in TOA uncertainty estimates, while EQUAD addresses additional sources of time-independent uncertainties, such as pulse jitter.

We used \texttt{TempoNest} \citep{2014MNRAS.437.3004L} to determine EFAC and EQUAD once we fitted the rotational parameters of the magnetar, maintaining the solution fixed and considering a small data span to discard red noise effects as described in \cite{zubieta_first_2023}.

\begin{figure}
    \centering
    \includegraphics[width=0.85\linewidth]{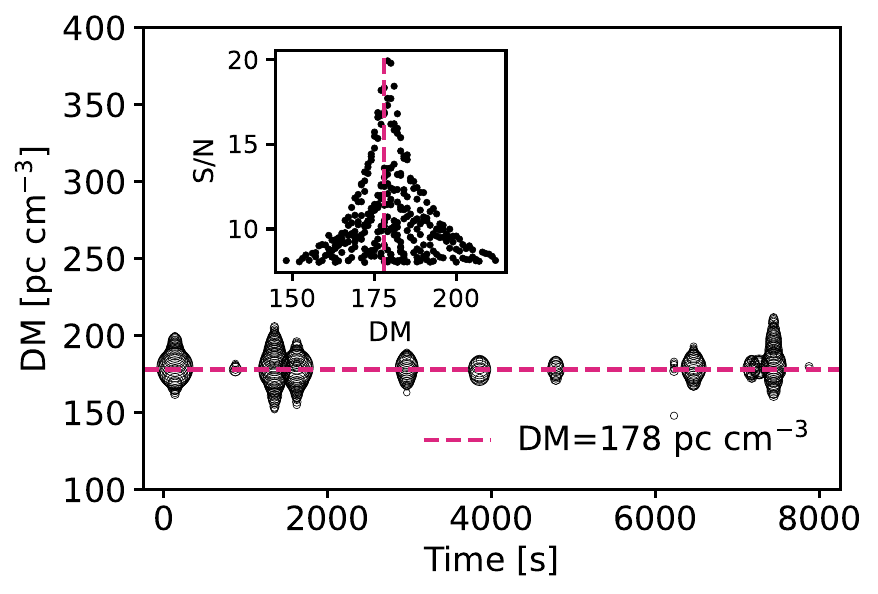}
    \caption{
    Single pulse search made with \textsc{PRESTO} on 19 November 2022 for the DM range. Time\,=\,0 corresponds to the start of the observation. 
    Each circle represents a pulse detected with a S/N\,$\geq$\,8, with its size proportional to its $\mathrm{S/N}$. 
    The dashed horizontal line marks the DM of the magnetar (178\dm).
    Pulses align in elongated islands centered around the magnetar DM. 
    Pulses that are farther from that line (as the one at $\sim$6200\,s) are more likely to be RFIs.
    Inset: distribution of the pulses' $\mathrm{S/N}$ as a function of DM. The pink vertical line marks the DM of the magnetar. The distribution of the pulses' $\mathrm{S/N}$ is also centered at the magnetar DM.}
    \label{fig:diagnosis_119}
\end{figure}

\subsubsection{Single pulses} \label{section:ReduccionSingle}

We studied the transient activity searching for single pulses. For this, we employed PRESTO \textsc{prepsubband} task to dedisperse the data. This produces a dedispersed temporal series at a DM value specified beforehand. We searched for single pulses between a DM range of 100-500 \dm, with incremental steps of 1 \dm. We did not use the parameter \textsc{zerodm} to help discard RFIs, as it produced artifacts on our time series that did not contribute to RFI classification. We searched each dedispersed time series for single pulses with PRESTO \textsc{single\_pulse\_search.py} script. We chose a threshold value of $\mathrm{S/N} >= 8$ \citep{maan_magnetar_2022} and the remaining parameters on default, searching for pulses in a downsampling range of 2 -- 30, or pulse width of 0.293 -- 4.389 ms.

\begin{figure}
    \centering
    \includegraphics[width=0.8\linewidth]{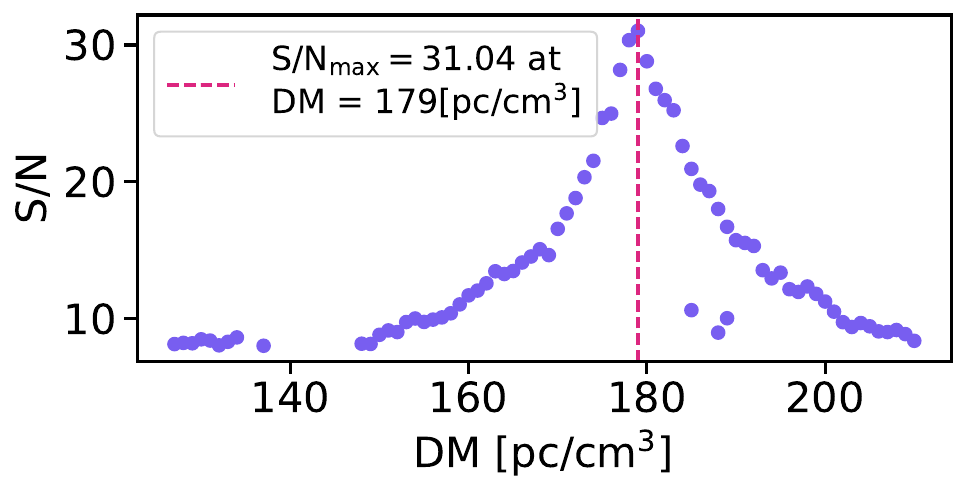}
    \caption{S/N vs DM distribution of one pulse detected on 25 November 2022, at time 7128.940~s from the start of the observation. It appeared at two ``different'' times: 7128.939\,s and 7128.940\,s. It was detected in 75 different time series across a DM range of 127--210\dm, peaking in S/N at $\mathrm{DM} = 179$\dm. }
    \label{fig:distribucion_confusin_pulso}
\end{figure}

We found that setting a threshold as high as $\textrm{S/N}\geq8$ was restrictive enough, as visual inspection of the diagnostics plots and the pulses themselves mainly revealed pulses from the magnetar. This is illustrated in a diagnostic plot from \textsc{single\_pulse\_search.py} in Fig.\,\ref{fig:diagnosis_119}, for 19 November 2022. The detected pulses appeared clustered around a DM of 178\dm. 

We extracted the candidates from all the searches and analyzed them with a Python-based script. Each circle in Fig.\,\ref{fig:diagnosis_119} represents a possible signal identified, hereafter referred to as an ``event''. A ``pulse'' is defined as a collection of events produced by the detection of a pulse across different time series (i.e., at varying DM values). To avoid multiple detections of the same pulse, we assumed that events at different DM values coinciding within a millisecond were produced by the same pulse. This millisecond threshold served as an initial empirical criterion, justified by the following: (1) the neutron star rotational period of $\sim$5.54~s implies that two pulses coincidental on the milliseconds should come from the same rotation, and (2) the widths of single pulses previously reported span a few milliseconds \citep{2022MNRAS.510.1996C, 2024A&A...686A.144B}. For each pulse, the DM value was taken from the highest S/N event within the group. This way, each group of circles depicted in Fig.~\ref{fig:diagnosis_119} was considered a single pulse if it met our criteria. 

Additionally, since previous studies estimated the DM of the magnetar within a range of 178--180\dm, we restricted our selection to pulses with DM values between 170--190\dm, resulting in a subset of 419 candidate pulses. This set included misclassified RFIs, duplicates, and real astrophysical pulses. We identified 249 single pulses through visual inspection of the diagnostic plots, as shown in Fig.\,\ref{fig:diagnosis_119}. The following step was to match the 249 visually identified pulses with the 419 candidates and to obtain a clean, definitive sample.

First, we fitted a Gaussian profile to each candidate. We defined the pulse width as the full width half maximum (FWHM), $\mathrm{FWHM}= 2\sqrt{2 \ln{2}} \ \sigma $, with $\sigma$ obtained from the fit. This parameter allowed a simple identification and RFI removal, as poorly fitted Gaussians produced outlier widths in the distribution. We manually cross-checked the remaining pulses, using both the diagnostic plots and the fitted profiles, and we discarded any remaining duplicates. High S/N pulses often appeared in several more time series than lower S/N ones. Due to our coincidence criterion ---based on truncating event times to the nearest millisecond--- some events from the same pulse were assigned slightly different timestamps, resulting in false duplication. For instance, a pulse occurring at 567.479956~s, and extending across several DMs could result in some events being truncated to 567.479~s and others to 567.480~s, and hence producing two identifications of the same pulse. In such cases, we checked whether the pulse centers were consistent and examined the S/N vs DM distribution of all events within $\pm 0.5$~s from the center ($\sim1/5$ of a period). Figure\,\ref{fig:distribucion_confusin_pulso} shows an example of a duplicated pulse. Through visual inspection of the time series, we confirmed the duplicates and assigned the correct DM based on the S/N peak. This process allowed us to reconstruct the properties of all 249 visually identified pulses.

\subsection{X-ray data}
\subsubsection{\textit{MAXI}}

We restricted our study of the light curve from the \textit{MAXI} data to the MJD range 59850--60428. We extended the analysis beyond the date range in radio after observing a peak in the light curve that recurred with a one-year interval. Following consultations with the \textit{MAXI} team, we had to distinguish whether the minor activity observed in the magnetar light curve was intrinsic to the source or was the result of contamination from the Sun or a nearby source.

\textit{MAXI} discards the observations of a source when it is close in the sky to the Sun within a certain angular distance. As the Sun was in an active phase within its 12-year activity cycle, contamination of the light curves due to solar leakage was still possible. The contamination can appear due to scattering of solar photons, mainly in the 2--4~keV band. Hence, a possible confusion arises as the magnetar activity is mostly seen in soft X-rays. The \textit{MAXI} team kindly performed an on-demand removal of solar contamination from the data. We worked with the cleaned light curves of \xte in both the 2--20~keV and the 4--20~keV energy bands. The difference between those light curves represents the activity in the 2--4~keV range. We cross-checked the light curves, the source region image, and the all-sky images available at \url{http://maxi.riken.jp/novasearch/} to see the daily variations and possible sources of contamination. In the all-sky images, the scatter from the Sun appeared as red diffuse halos originating from the point where the Sun is located. 

Another potential contamination in the source light curves could be caused by nearby X-ray sources, primarily GX 9+1 and GX 13+1. Due to their small angular distance to \xte, there was a chance that the observed variability could have originated from the intrinsic variability of such neighboring X-ray binaries. The \textit{MAXI} team decontaminated the light curves by extending the neighboring source regions to reduce contamination. Additionally, if the elongated point spread function (PSF) of the \textit{MAXI} instrument experienced any distortion or change in its position angle, this could also produce contamination. We examined the light curves of GX 9+1 and GX 13+1 simultaneously with the light curves of the magnetar. For the days on which observations of the magnetar and the other sources were available, we computed the correlation of the X-ray flux in the 2--20~keV band. We used the \texttt{corr} method in the \texttt{pandas} library and calculated the Pearson correlation coefficient. Finally, we analyzed the presence of the magnetar emission and the PSFs of both GX sources in the images, considering a FoV of $\sim 15^{\circ} $, combined with the variations in the light curve. In section \ref{sec:res_maxi} we explore in detail when contamination could be ruled out and when it could not.

\begin{figure}[!ht]
\centering
 \includegraphics[width=1\linewidth]{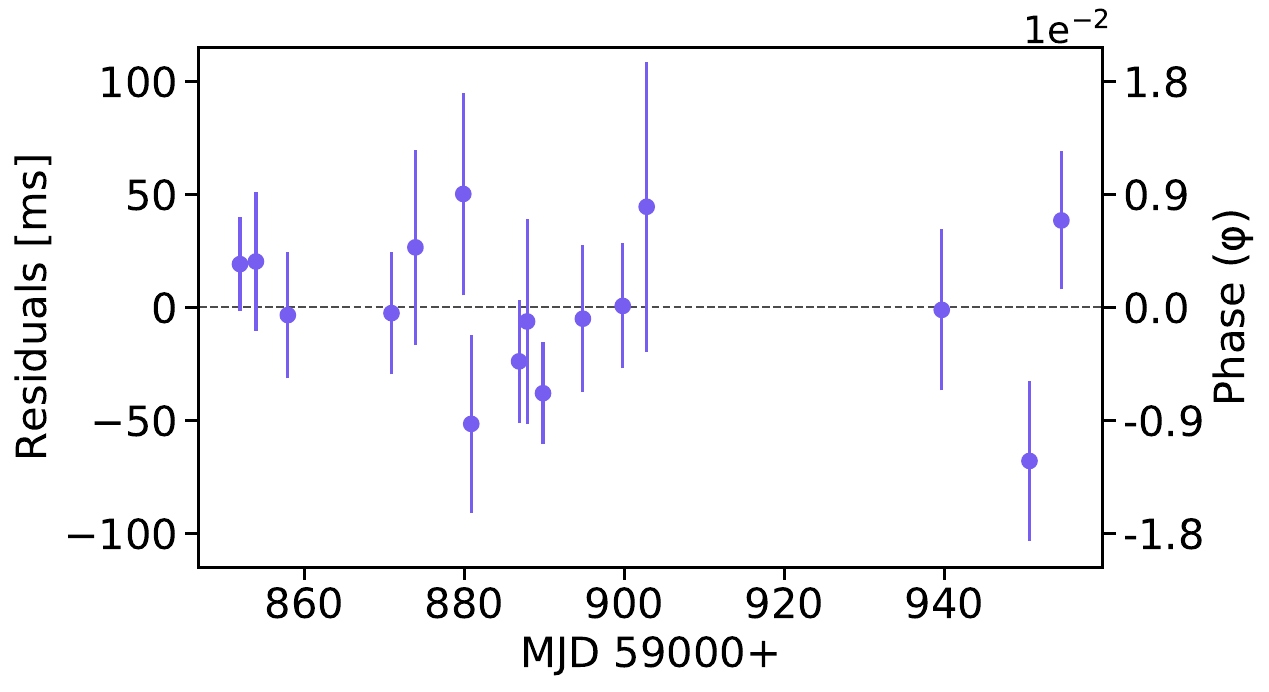}
   \caption{Residuals with respect to the timing model presented in Table \ref{tab: timing-solution}.}
  \label{fig:residuals}
\end{figure}

\section{Radio results}

\subsection{Timing analysis}

We monitored the pulsar from 29 September 2022 (MJD 59851), until 14 July 2023 (MJD 60140). However, we only detected the folded profile until 10 January 2023 (MJD 59954).  Within this date range, we performed 60 observations of the magnetar and detected it 16 times.

\begin{table}[htbp]
    \caption[]{Results for the timing model by fitting to the TOAs analyzed in this work.}
    \centering
    \label{tab: timing-solution}
    \begin{tabular}{lc}
        \hline
        \noalign{\smallskip}
        Parameter & Value \\
        \noalign{\smallskip}
        \hline
        \noalign{\smallskip}
        $t_0~[\mathrm{MJD}$] & $59849$ \\
        $\nu~[\mathrm{Hz}]$ & $0.180420042(9)$ \\
        $\dot\nu~[\mathrm{Hz~s^{-1}}]$ & $-2.69(5)\times10^{-13}$ \\
        $\ddot\nu~[\mathrm{Hz~s^{-2}}]$ & $5(1)\times10^{-21}$ \\
        $\mathrm{DM^\dag}~[\mathrm{pc~cm^-3}]$ & 178 \\
        $\mathrm{EFAC}$ & 0.7(1) \\
        $\mathrm{EQUAD}$ & -7.4(14) \\
        \noalign{\smallskip}
        \hline
    \end{tabular}
    \tablefoot{$^\dag$ Extracted from \cite{2022MNRAS.510.1996C}.}
\end{table}

To obtain the timing solution for these observations, we fitted $\nu$, $\dot \nu$ and $\ddot \nu$ to the TOAs, keeping DM fixed at 178\dm\citep{2022MNRAS.510.1996C}. Results are shown in Table \ref{tab: timing-solution} and the residuals for the model are shown in Fig.~\ref{fig:residuals}.
We obtained a phase-coherent solution between MJD 59851 and MJD 59954. After this epoch, we were not able to detect the folded profile of the magnetar. This reduction in the activity of the pulsar agrees with the diminishment of the activity of GPs.

\subsection{Single pulses}
\subsubsection{Transient activity}

Over 149 days of observations, we detected single pulses in 56 days. In the remaining 93 days we were not able to detect pulses with our criteria.

\begin{figure}[!ht]
    \centering
    \includegraphics[width=1
\linewidth]{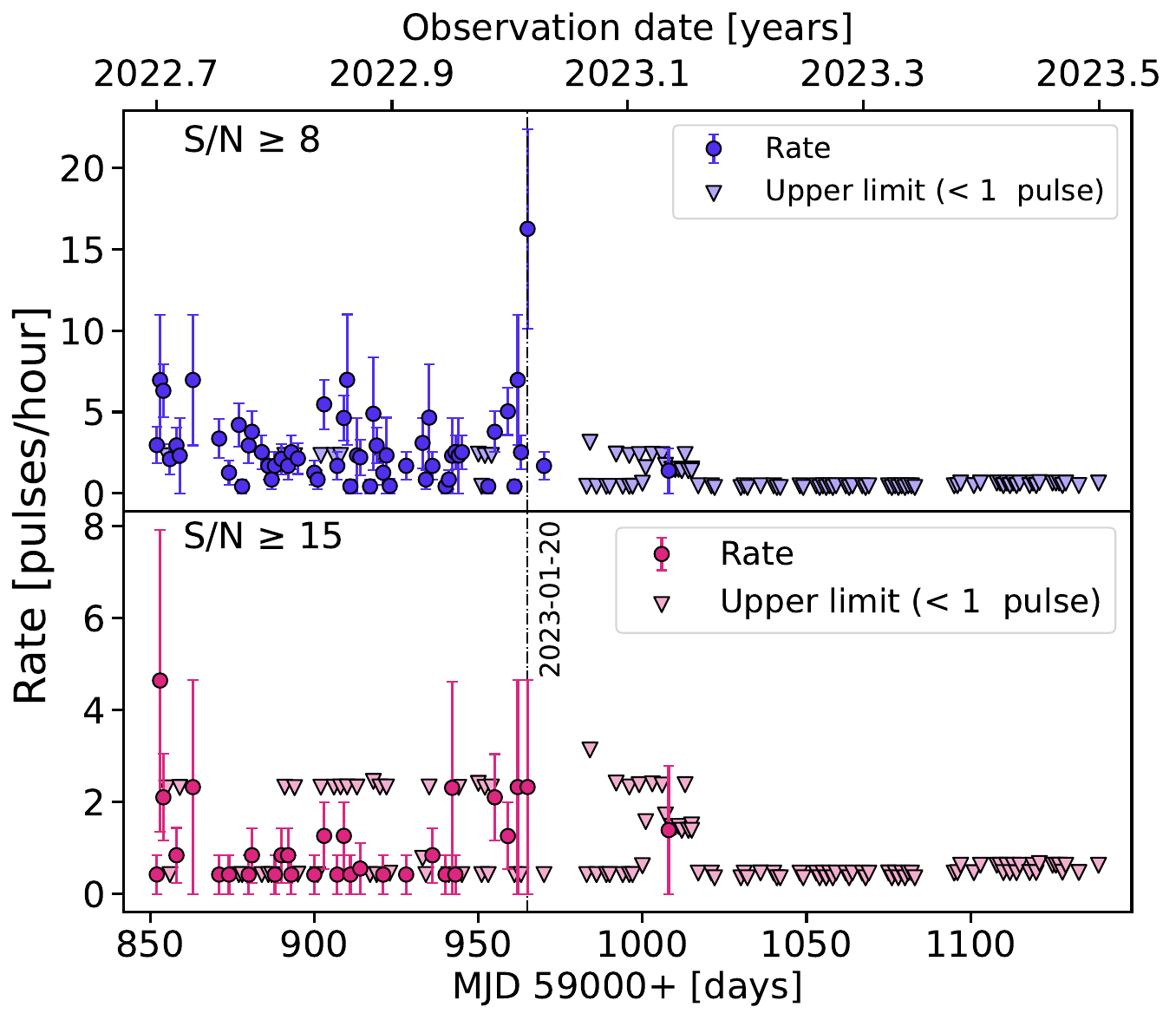}
    \caption{Rate of pulses per hour of observation against days of observation. The upper panel shows the rate considering only pulses of $\mathrm{S/N} \geq8$ and the lower panel only of $\mathrm{S/N}\geq15$. In circles we represent the rate and in filled triangles the upper limits. }
    \label{fig:rate}
\end{figure}

The final sample consists of 249 single pulses from the magnetar. We analyzed the transient activity in terms of the rate, i.e., the number of pulses per hour. To obtain the rate, we determined the effective observing time $t_{\mathrm{observed}}$ for each observation by taking the total exposure time and excluding intervals in which more than 70\% of the frequency channels were masked. 

We assumed the emission process to be Poissonian, and we assigned to the number of pulses from a day, $N_{\mathrm{day}}$, an uncertainty of $\sqrt{N_{\mathrm{day}}}$. Finally, we calculated the rate as $N_{\mathrm{day}}/t_{\mathrm{observed}}$. For non-detection days, we estimated an upper limit as $1/t_{\mathrm{observed}}$ indicating that $<1$ single pulses were detected on those days, following the method of \cite{lazarus_constraining_2011}. Since the observation durations ranged from 18~min to 172~min, shorter observations on non-detection days result in higher upper limits. 

The resulting rate is shown in Fig.~\ref{fig:rate}. In the upper panel, we show the rate considering all the single pulses.  A peak in activity occurred on 20 January 2023, after which only two days had detected single pulses. This suggests that the magnetar experienced a highly active state before its radio emission dropped below the sensitivity limit of the IAR telescopes. In the lower panel, we present the rate only considering pulses with S/N$\geq$15, to highlight the behavior of the most energetic pulses. In this case, the upper limit represents non-detection of pulses with S/N $\geq$15. This decomposition reveals that the rate of more energetic pulses was higher earlier in the campaign, despite the overall peak in activity occurred shortly before the subsequent decline. We computed the minimum detectable flux density of single pulses with the next relation (e.g., \cite{pastor-marazuela_new_2023}): 

\begin{equation}
    S_{\mathrm{lim, sp}} = (S/N)_{\mathrm{min}} \frac{\beta \  T_{\mathrm{sys}}}{G\sqrt{n_\mathrm{p}\ \Delta\nu \ t_\mathrm{obs}}} \ \sqrt{\frac{t_\mathrm{obs}}{W}},
\end{equation}
where $G$ in K/Jy is the telescope gain, $\beta\sim1$ the digitazion factor for the IAR,  $t_\mathrm{obs}$ the observation time and $W$ is the width considered, the remaining values are the same from Equation \ref{eq:gain_tnoise}. In general, when there is no detection, a broad value of $W$ is used, such as 10~ms. In this case, we considered a value of 4.4 ms as it was the broader width in our search. The result is a flux density and luminosity upper limits of $S_{\rm{min}} = 7\pm1$~Jy and $L_{\rm{min}}=600\pm100$~Jy\,pc$^{2}$ for a single pulse.

\begin{figure}[!ht]
\centering
 \includegraphics[width=0.9\columnwidth]{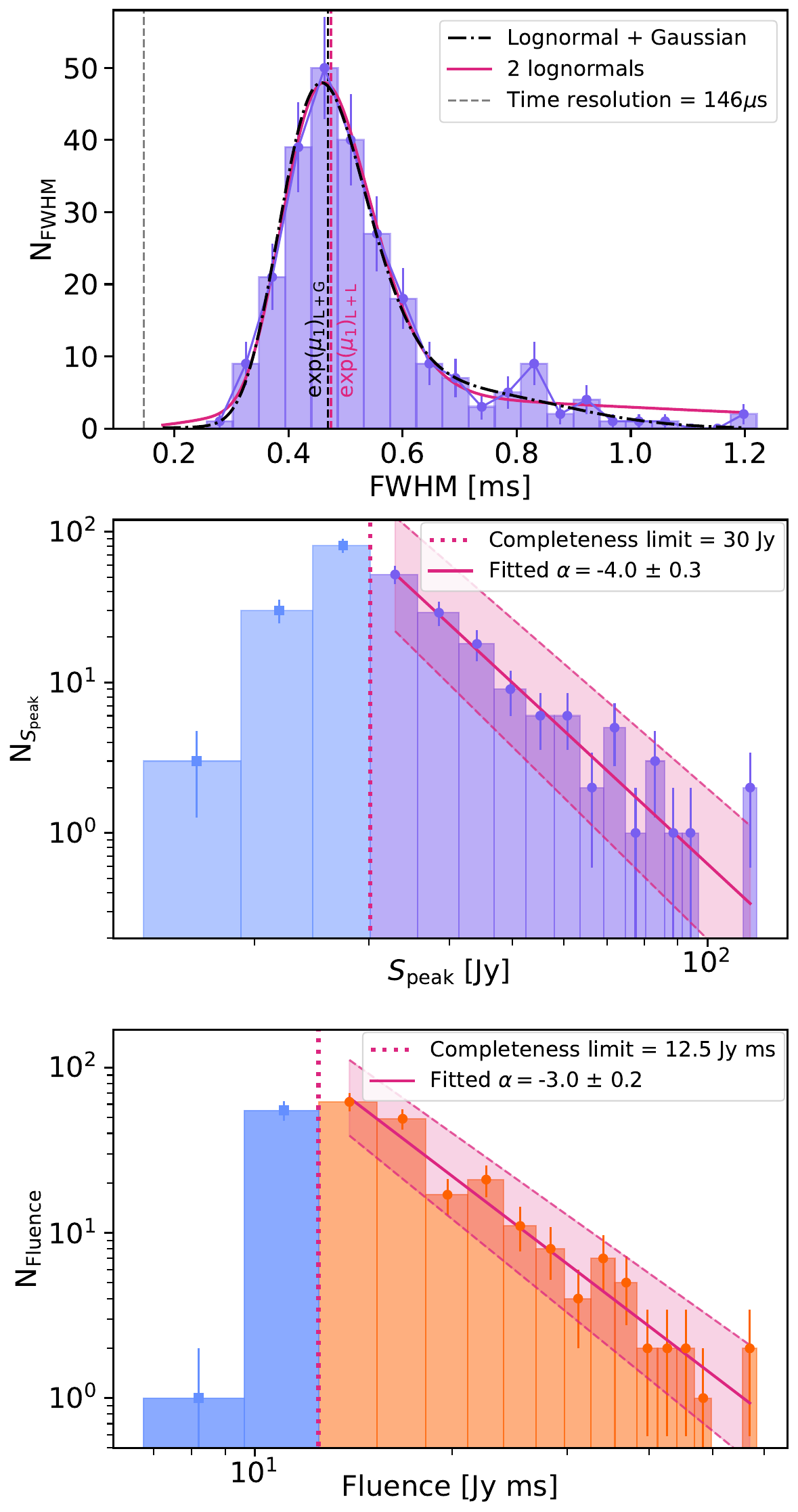}
   \caption{Histograms of pulses grouped by different properties, together with the best-fit of the distribution. \\ \textit{Upper Panel:} FWHM distribution fitted fitted with L+G (black dotted) and L+L (pink continuous curve) models. Vertical lines show: the time resolution (gray), the $\mathrm{exp}(\mu_{1})$ value of the lognormal component of the L+G model (black) and the $\mathrm{exp}(\mu_{1})$ value of the first lognormal component of the L+L model (pink). \\ \textit{Middle Panel and lower panel :} flux density and fluence distributions with the respective completeness limit indicated by vertical dotted lines, the sloped line indicates the fit and the shadowed area the uncertainty range.}

  \label{fig:flux_curvefit}
\end{figure}

\subsubsection{Pulses properties} \label{sec:ajustes}
We analyzed the distribution of the width of the pulses (FWHM), flux density, and the fluence (or energy) of all single pulses. We consider that all of our single pulses are giant pulses (see Section \ref{sec:onlyGP}).

\begin{table}[htbp]
\centering
\begin{tabular}{c c c}

\hline
        \noalign{\smallskip}
{Parameter} & {L+G} & {L+L}  \\
\noalign{\smallskip}
        \hline
        \noalign{\smallskip}
$A_{1}$  & 8.9 $\pm$ 1.5 & 8.6 $\pm$ 0.9  \\
$\mathrm{exp}(\mu_{1})$  & 0.469 $\pm$ 0.008 & 0.474 $\pm$ 0.004 \\
$\sigma_{1}$  & 0.17 $\pm$ 0.02 & 0.17 $\pm$ 0.01 \\
$A_{2}$  & 5 $\pm$ 2 & 5 $\pm$ 3 \\
$\mu_{2}$ or  $\mathrm{exp}(\mu_{2})$ & 0.70 $\pm$ 0.13 & 0.9 $\pm$ 0.5 \\
$\sigma_{2}$ & 0.18 $\pm$ 0.06 & 0.6 $\pm$ 0.3  \\

 \noalign{\smallskip}
        \hline
         \noalign{\smallskip}

d.o.f. & 13 & 13  \\
 \noalign{\smallskip}

$\chi^2_{\mathrm{red}}$ & 0.78 & 1.42\\

 \noalign{\smallskip}
        \hline
\end{tabular}
\caption{Parameters for the fitted models of the FWHM, with lognormals (L) and gaussian (G). In the case of the L+G model, the parameter for the gaussian is $\mu_{2}$, and for the L+L model, the parameter of the second lognormal is $\mathrm{exp}(\mu_{2})$}
\label{tab:model_combinations}
\end{table}

We fitted the FWHM distribution using \texttt
{curvefit} from \texttt{scipy} with two different models. One consisted of a lognormal plus a Gaussian distribution (L+G) and the other of two lognormal functions (L+L). The upper panel of Fig.~\ref{fig:flux_curvefit} displays the resulting distributions overlaid on the data. There is a principal component in the range of 0.2 to 0.7~ms, and a second component for higher FWHM. In Table \ref{tab:model_combinations}, the resulting parameters are shown, along with the $\chi^2_{\mathrm{red}}$ and the number of degrees of freedom (d.o.f.). The L+G model is favored, with a $\chi^2_{\mathrm{red}}=0.78$ against one of $1.42$.

Previous studies of the flux density distribution of the pulses from the magnetar have modeled it as a power-law function \citep{maan_distinct_2019} or a broken power law \citep{2022MNRAS.510.1996C}. To study our distribution, we applied a flux density threshold at 30~Jy, at the transition of the peak of the distribution, this limit is consistent with the lower bound of the power-law obtained in the fit (the parameter \texttt{x min}). We studied the distributions in two ways. First, we studied $N_{S_{\mathrm{peak}}}$, the number of pulses detected in a 5.5~Jy flux density interval, using a Knuth binning in the linear space.  We assume the emission process is Poissonian and assign error bars as the square root of the number of pulses in each bin. We employed \textsc{curve\_fit} from \textsc{scipy}, which uses a nonlinear least squares fitting of our model, a power-law $N_{S_{\mathrm{peak}}} \propto S_{\mathrm{peak}}^{\alpha}$.  The optimized value obtained was $\alpha = -4.0\pm0.3$. In the middle panel of Fig.\,\ref{fig:flux_curvefit}, we show our full sample and indicate the completeness limit. We plot the distribution on a logarithmic scale, where the power-law behavior is represented with straight lines. The different colored bins represent which bins were used for the fitting and which were ignored.

The second method comprised a power-law fit to the flux density's Probability Density Function (PDF). We used the Python package \textsc{powerlaw} \citep{alstott_powerlaw_2014} to obtain the power-law index ($\alpha$) that best fits the data. This method employs Pareto's maximum-likelihood estimate (MLE) to determine the optimal index. It gives the minimum flux density value (\texttt{x min}) to adjust the power law by finding the minimal Kolmogorov-Smirnov distances between the power law fitted and the data. In this case, it was 30.18~Jy, which is our completeness threshold. We obtained the uncertainty of $\alpha$ by performing a bootstrap with replacement, 1000 times. We obtained $\alpha=-4.1\pm0.4$ for the distribution. Since both fitted values agree, we use the first method's value; the corresponding fit appears in the middle panel of Fig.\,\ref{fig:flux_curvefit}.

\begin{figure}
    \centering
    \includegraphics[width=0.85\linewidth]{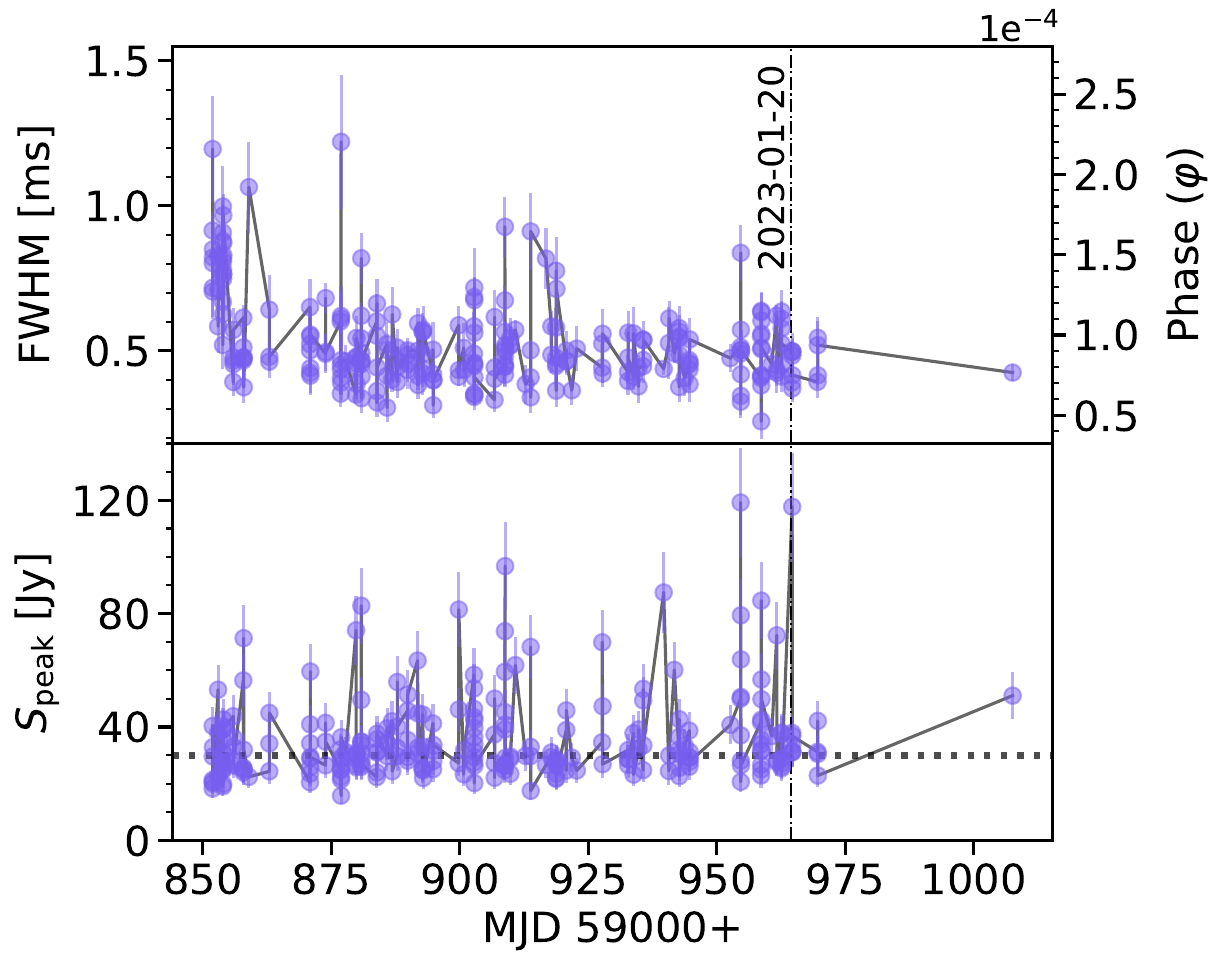}
    \caption{FWHM and flux density variation through the days. In the lower panel, the dotted horizontal line is the flux density completeness limit.  }
    \label{fig:variacion_ancho_y}
\end{figure}  

\begin{figure}
    \centering
    \includegraphics[width=0.85    \linewidth]{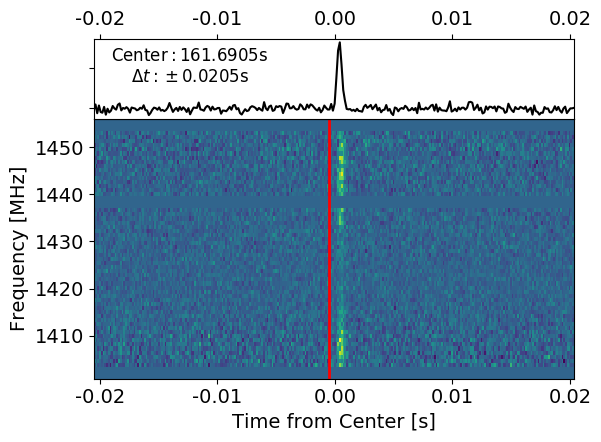}

    \caption{Dynamic spectrum (lower) and time series (upper) for the brightest pulse of 20 January 2023, dedispersed at 179\dm. The center of the time series (161.6905~s) represents the time from the start of the observation. The vertical red line is shown for reference and corresponds to a pulse of $\mathrm{DM}=179$\dm. }
    \label{fig:waterfaller}
\end{figure}

\begin{figure}[htbp]
    \centering
    \includegraphics[width=0.7\linewidth]{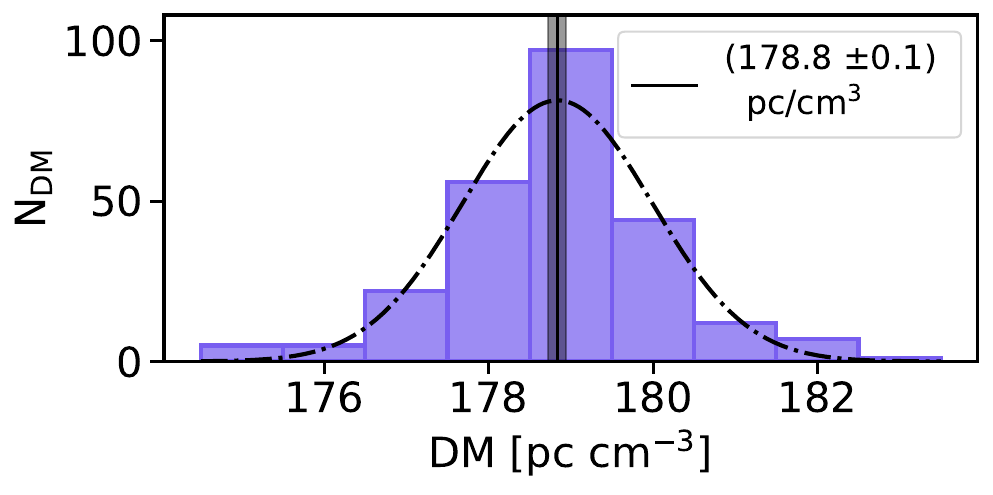}\\
    \includegraphics[width=0.7\linewidth]{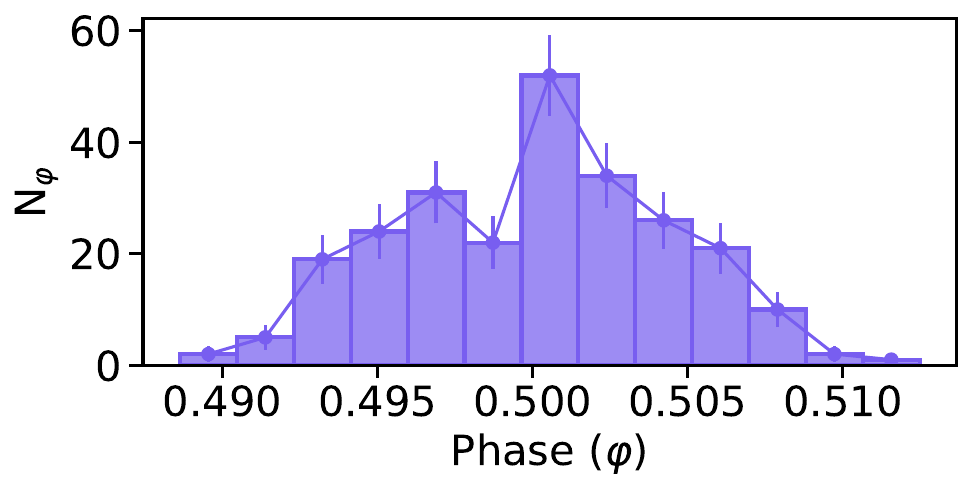}\\
    \includegraphics[width=0.7\linewidth]{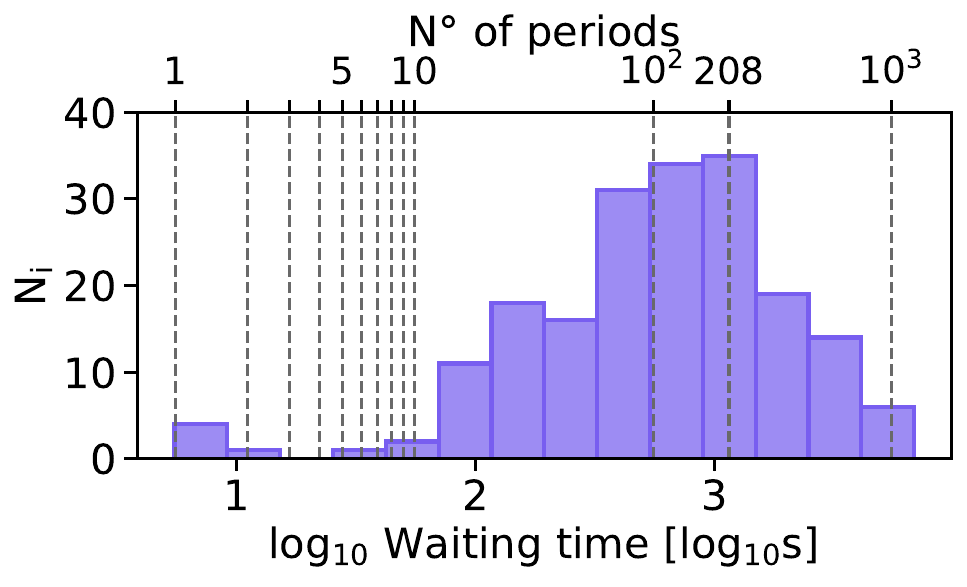}
    \caption{\textit{Upper panel}: distribution of the pulse dispersion measure of the pulses, the dash-dotted line is the Gaussian fit. The vertical line is $\mu$ and the colored area represents the uncertainty of the value. \textit{Middle panel}: distribution of pulse phases. \textit{Lower panel}: distribution of $\mathrm{log_{10}}$ of the waiting time between two consecutive pulses. The vertical lines indicate waiting times equal to multiples of the spin period.}
    \label{fig:distribuciones_vertical}
\end{figure}

For the fluence, we applied the same methodology as for the flux density. We considered a fluence completeness limit of 12.5\fluence, obtained empirically and confirmed with \textsc{powerlaw}. In this case, we studied the number of pulses $N_{\rm Fluence}$ in a 2.9\fluence interval. The index of the fitted power-law with \textsc{Scipy} was
$\alpha = -3.0\pm0.2$ and with \textsc{powerlaw} $\alpha = -3.3\pm0.5$. As before, we considered the value from the first method.

Figure\,\ref{fig:variacion_ancho_y} presents the time evolution of both the FWHM and the flux density. The FWHM appears to decrease and stabilize over time. The vertical line marks the day with the highest rate of detected single pulses, coinciding with one of the brightest single pulses with $S_{\nu}=118\pm19\mathrm{\,Jy}$. The brightest pulse was detected on 10 January 2022 (MJD 59954), with a flux density of $119\pm19 \mathrm{\,Jy}$, while the faintest detected pulse occurred with $S_{\nu}=16\pm3 \mathrm{\,Jy}$ on 24 October 2022. Following this peak activity ---marked by both the highest pulse rate and brightest pulses--- the magnetar's activity diminished substantially, becoming virtually undetectable for the IAR sensitivity.

\subsubsection{Morphology, dispersion measure, phase, and periodicity}

The sample of single pulses shows no particular behavior on their dynamic spectrum. In Fig.\,\ref{fig:waterfaller} we dedispersed the brightest pulse on 20 January 2023, at 179\dm. The diminished central band intensity aligns with the known frequency response characteristic of our receiver.

As previously mentioned, the pulses had a range of possible DM values. When producing plots as shown in Fig.\,\ref{fig:waterfaller}, we confirmed the astrophysical origin of the pulses. Our DM resolution, based on our sample time, was approximately $0.93$\dm. Technically, our uncertainty in each measured DM is of 1 \dm. However, by analyzing the pulses distribution, we can statistically narrow down the magnetar DM. We fitted a Gaussian distribution to the number of pulses per DM (upper panel of Fig.~\ref{fig:distribuciones_vertical}). The parameters were $\mu=178.8\pm0.1$\dm, $\sigma=1.15\pm0.09$\dm, with a $\chi^2_{\mathrm{red}}=2.26$ and 6~d.o.f.

We studied the dispersion in the phase of all our GPs, as shown in the middle panel of Fig.\,\ref{fig:distribuciones_vertical}. For each date, we calculated the period with our timing solution and assumed it was constant throughout the observation. The maximum phase difference for pulses within the same observation was $\Delta\varphi=0.0196$, which represents approximately $2\%$ of the magnetar's rotational period. Since we had days with detected GPs but not an integrated pulse profile, this prevented us from consistently aligning the GPs based on a timing solution. Thus, we aligned the mean phase of each day by manually shifting each to 0.5, and applied the same shift to each individual phase. The maximum difference in shifted phases across all observations was $\Delta \varphi=0.0239$, which represents approximately $ 2.4\%$ of the neutron star's rotational period. Our results indicate that the detected emission originates from roughly $2.4\%$ of the magnetar's rotational phase. This small phase range corresponds to an observed opening angle of the emission region of $8.64^{\circ}$. We found no correlation between pulse phase and brightness ($S_{\nu}$), with Pearson and Spearman correlation coefficients of 0.03 and 0.02, respectively. 

We analyzed the waiting-time distribution between consecutive single pulses to determine the typical interval required for subsequent pulse detection. The lower panel of Fig.\,\ref{fig:distribuciones_vertical} displays this distribution in both logarithmic units ($\mathrm{log_{10}(waiting\ time)}$) and in units of magnetar rotation periods. We only observed one case of two consecutive rotations with GPs detected. The waiting time shows a main distribution center around 100--208 periods and a secondary minimum between 1--3 periods. A waiting time shorter than one period would indicate multiple peaks in one single rotation, i.e., the detection of substructure. The observed distribution is evidence of the non-detection of substructure in our observations. The mean waiting time is 562.7~s (101 periods) with a standard deviation of 3.7~s (0.7 periods). 
\begin{figure}
    \centering
     \includegraphics[width=0.49\textwidth]{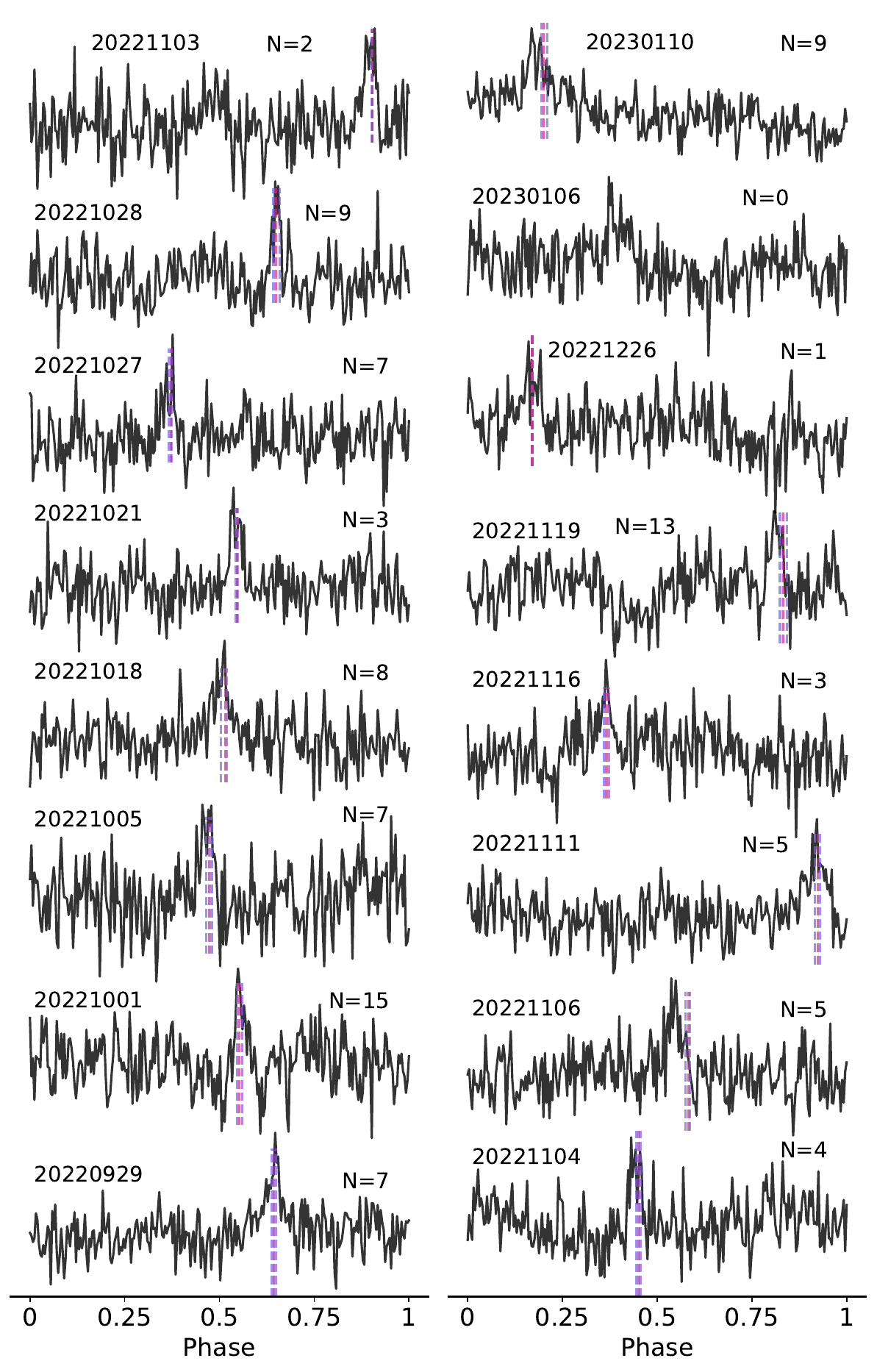}

    \caption{Integrated pulse profiles for the days with detection. For each day, the number of single pulses detected is indicated by N. Vertical lines indicate the mean phase (in pink) along with the minimum and maximum pulse phase values (in violet) for individual pulses. }
    \label{fig:profiles}
\end{figure}

Previous studies of this magnetar have shown that single pulses and giant micro pulses are emitted across a wide range of rotational phases, generally coinciding with the window of the integrated pulse profile \citep{kramer_polarized_2007, serylak_simultaneous_2009, dai_wideband_2019,maan_distinct_2019, \bause}. In contrast, GPs appears at a limited phase range  \citep{serylak_simultaneous_2009,2022MNRAS.510.1996C}, with GPs from the 2003 outburst limited to $\sim4^{\circ}$ \citep{serylak_simultaneous_2009} and for the 2018 in a range of $\sim9-10^{\circ}$, either preceding or lagging the folded profile \citep{2022MNRAS.510.1996C}. Figure\,\ref{fig:profiles} displays the 16 integrated pulse profiles and their corresponding single pulse phase range, indicated by vertical dotted lines. For pre-November 2022 observations, the single-pulse phase range aligns with the peak of the noisy integrated profile. From November onward, we observe a trailing phase drift in some observations, most notably in the profile from 6 November 2022. Of the 16 days with detected profiles, only one lacked detection of single pulses. Every day showing both profiles and single pulses contains at least one pulse with S/N $\geq 10$, including two days with a single GP detected (S/N = 13 and 28). Of 78 days with GPs but without profiles, many exhibit pulses with S/N  $\geq 10$, particularly the bright GP on 20 January 2023 (S/N = 40, $S_{\nu}= 118 \pm 19 \mathrm{\,Jy}$). 

These results align with prior studies \citep{serylak_simultaneous_2009,2022MNRAS.510.1996C, 2024A&A...686A.144B}, suggesting the emission comprises two distinct components: a 'normal' (steady) component and a GP component. On the days with detected profiles, the profile phase width is wider than the single-pulse phase range. This difference may arise from the underlying 'normal' component or from single pulses below our instrumental sensitivity. This is shown by days of non-alignment between the profile peaks and single pulse phase range, and by one profile detection without corresponding single pulses. 

\section{X-ray results: MAXI} \label{sec:res_maxi}

We examined the magnetar's light curve to check whether the detected peaks at around 0.1-0.2 c/s/$\textrm{cm}^{2}$ at 2--20 keV were produced by its activity or due to contamination. The peaks that recurred with a one-year interval disappeared in the Sun-cleaned data. However, some smaller peaks persisted on other days. We paid particular attention to three date ranges, namely:  29 September to 9 December 2022 (MJD 59851--59922),  2 January to 5 June 2023 (MJD 59946--60100), and  13 September to 4 December 2023(MJD 60200--60282), as these periods exhibited increases in the measured flux. We found that some days appeared to be contaminated by diffuse scattering from the Sun, but not all. 

When we analyzed the light curves of the GX sources, they did not exhibit the same behavior as \xte. The Pearson correlation coefficient indicated no correlation between their fluxes with that of the magnetar, with values of $-0.12$ for GX 9+1 and $0.14$ for GX 13+1. Although these values suggest a lack of correlation, further examination of the PSF distortion revealed that the peaks observed in the light curve of the magnetar tended to occur when the PSFs of the other sources were elongated and aligned with the position of the magnetar. The shape and orientation of the PSFs varied from day to day, affecting or producing the peaks detected in the magnetar light curve.

We found that all variations in the X-ray light curves can be attributed either to contamination from the Sun or to a distorted PSF from neighboring bright sources. On some days, both sources of contamination were present. The magnetar itself did not exhibit any detectable activity during this period in the data collected by the \textit{MAXI} instrument.

\section{Discussion}

\subsection{Giant pulses}\label{sec:onlyGP}
Our sample consists exclusively of giant pulses. However, as we have mentioned, there is no universal consensus about how a GP should be defined. Various studies adopt different criteria: some define a GP when its energy exceeds ten times the average energy of all single pulses; others use a similar characterization but invoking flux density; and yet others if its flux density is ten times the average peak flux density of the folded profile\footnote{The adoption of the number ten in these definitions suggests it is a community-adopted threshold for distinguishing between populations.}. In previous studies of this magnetar, \cite{2022MNRAS.510.1996C} reported GPs with peak flux densities up to 158~mJy. In this work, we detected pulses with flux densities from $16\pm3$~Jy up to $119\pm19$~Jy, i.e., up to three orders of magnitude higher in peak flux density. Based on this comparison, we consider that the entire sample of pulses consists only of giant pulses. This type of criterion appears in literature ---especially in studies with the Crab pulsar, for instance \cite{karuppusamy_giant_2010,doskoch_statistical_2024}--- where a pulse is considered a GP if its S/N exceeds a certain threshold. 

Another criterion often used to identify GPs is the study of the flux and fluence distribution. Typically, GPs describe a power-law distribution, in contrast with the normal or lognormal distributions of normal pulsar emission. We performed this analysis on the IAR data and fitted a power-law distribution to the flux density and fluence of the pulses, supporting the GP classification. Additional evidence comes from the narrow phase range reported, especially when compared to the broader apparent phase width of the integrated profiles in Fig.\,\ref{fig:profiles}.

A comparison of the flux density and fluence distributions with other works is as follows. For the previous outburst, \cite{serylak_simultaneous_2009} reported a power-law index of -1.85 for the energy distribution of GP at higher frequencies (8.5 GHz). Our results indicate a steeper energy distribution for the 2018 outburst. \cite{2022MNRAS.510.1996C} reported a broken power law for the flux density with a power of about $-1$ or $ -4$, depending on the flux density range. \cite{maan_distinct_2019} found that the high-energy tails of the distribution, where the brightest pulses showed similarities to giant pulses, were described by a simple power-law with a mean index of $\sim -3.3$ across five observations. \cite{2024A&A...686A.144B} noted that their single pulses did not follow a power-law, possibly due to the different definition of a single pulse used. Following the behavior observed in the previous outburst, where pulse energy distributions changed daily, our sample might indicate an evolution between the epoch studied by \cite{2024A&A...686A.144B} and our observations.

\begin{figure}
    \centering
    \includegraphics[width=0.9\linewidth]{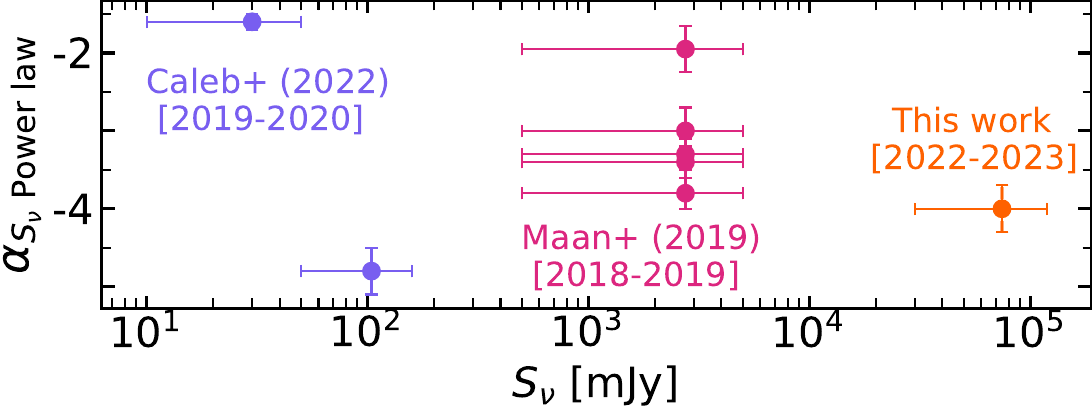}
    \caption{Power law index fitted by different studies for GP and giant micro pulses after the 2018 outburst, color-coded by the reference and with the observation years indicated between brackets.}
    \label{fig:index}
\end{figure}

We fitted a single power-law of index $\alpha = -4.0\pm0.3$ to the flux density distribution, and $\alpha = -3\pm0.2$ to the fluence distribution. It is important to note that the flux density range and the epoch at which each study derived these indices differ, with no overlap in the flux density intervals, as shown in Fig.\,\ref{fig:index}. This work corresponds to the highest flux density range. The differing values could indicate a dependence on the $S_{\nu}$ range considered or the employed bandwidth, as \cite{maan_distinct_2019} observed at lower frequencies, but primarily reflect the evolving nature of the emission. Previous studies on GPs for the Crab pulsar found a power law of index about $ -3$ for its flux density distribution \citep{karuppusamy_giant_2010}, the index obtained in this work is steeper.

\subsection{Some implications of our observations}
Following its outburst, the magnetar entered a period of mild transient activity in December 2018. Two subsequent peaks in radio activity were observed before our campaign ---one in September 2020 and another in February 2021. The September 2020 peak marked the onset of a more sustained active phase \citep{2022MNRAS.510.1996C,\bause}. The IAR observations began in September 2022 and revealed enhanced activity by February 2023, followed by a sharp decline below our sensitivity limit. This suggests that the peak in radio emission occurs significantly delayed with respect to the initial X-ray outburst. 

Previous efforts to identify a correlation between X-ray and radio activity yielded no results.  \cite{2022MNRAS.510.1996C} reported no X-ray enhancement accompanying the radio activity. They even found an X-ray flux minimum coinciding with the rise in average radio flux density toward September 2020. Similarly, \cite{\bause}  found no associated X-ray activity in archival \textit{Swift}/XRT data from August 2020 and February–April 2021. In our analysis, we found no reportable activity on the magnetar from the \textit{MAXI} observations.

The measured widths are the narrowest reported to date. We note two caveats regarding observed pulse width comparisons. First, the different methods employed make a direct comparison difficult. Second, our flux density sensitivity of $S_{\mathrm{min}}=7\pm1$~Jy (for S/N=8) means that a more sensitive telescope would likely measure wider FWHM values for the same pulses. While previous studies have detected pulses several orders of magnitude fainter, our faintest pulse is $16\pm3$~Jy. This empirical detection threshold is higher than the theoretical sensitivity calculated for the ETTUS boards in Section 5.2.1. This suggests we are likely detecting only the brightest portion of each pulse. If the pulses follow a Gaussian profile, we would primarily observe the narrow peak region, explaining the small measured widths. 

For a resolved single pulse with a known width ($\Delta t$), the size of the emission region can be estimated using the light travel time: $r <c\Delta t$ \citep{2004hpa..book.....L}, where $c$ is the speed of light. Considering the average FWHM measured in this study, $0.469\pm0.008 \mathrm{\, ms}$, we estimate a size of the emitting region of $141\pm 2$~km. If the intrinsic width is broader, e.g., 1--10~ms, as reported in previous studies at similar frequencies, the corresponding size would increase to $\sim$300--3000~km. In any case, the radiation seems to be produced well within the magnetosphere (the radius of the light cylinder in XTE~J1810-197 is $\approx 2.65 \times 10^{10}$ cm). 

The single pulse luminosities range from $(6\pm2)\times10^{30}\mathrm{\, erg \, s^{-1}}$ to $(5\pm2)\times10^{31}\mathrm{\, erg \, s^{-1}}$, with an average value of  $(1.5\pm0.7) \times 10^{31}\mathrm{\, erg \, s^{-1}}$, approximately $10^{-2}$ times lower than the magnetar's spin-down luminosity of $4.6\times10^{33}\mathrm{\, erg \, s^{-1}}$, as derived from our timing solution and previously reported $B_\mathrm{surf}$ values. The brightness temperatures range from $(3\pm2)\times 10^{25}\mathrm{\, K}$ to $(5\pm2)\times 10^{26}\mathrm{\, K}$, with a mean value of $(5\pm3 )\times 10^{26}\mathrm{\, K}$, about a factor $10^{-6}$ lower than what was reported for FRB20200426 \citep{bochenek_fast_2020} at a similar central frequency. All these temperatures, however, are well above the Compton limit of $10^{12}$ K, indicating that the radiation must be coherent.

\begin{figure}
    \centering
    \includegraphics[width=1\linewidth]{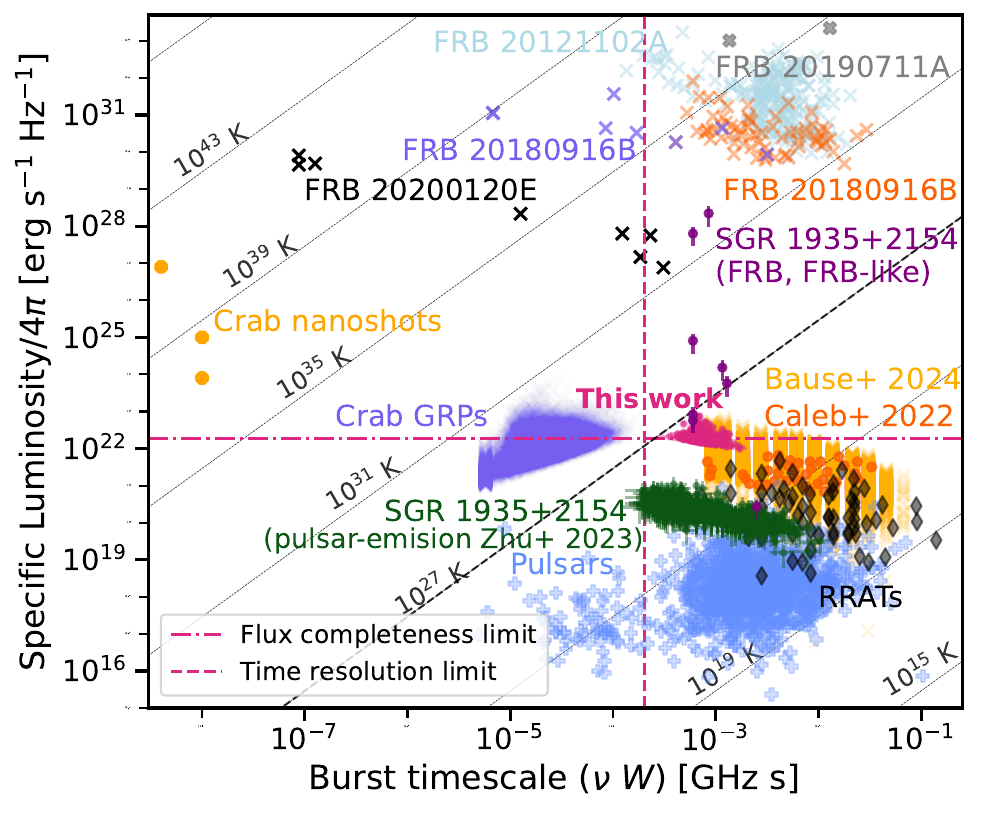}
    \caption{Transient phase space adapted from \cite{nimmo_burst_2022} including the GPs detected in this work and their error bars, previous works on \xte \citep{2022MNRAS.510.1996C, 2024A&A...686A.144B}, and a recent work on FAST radio pulses from SGR\,J1935+2154 \citep{zhu_radio_2023}. In the case of SGR\,1935+2154, the vertical error bar of each pulse includes the uncertainty in the magnetar distance. The pink horizontal dotted line represents our flux completeness limit, and the vertical one our time resolution limit.}
    \label{fig:trans_spacs}
\end{figure}

\subsection{Is it possible to observe an FRB from this source?}\label{sec:FRBquizas}

An important question about this source is whether it could present a similar behavior to SGR\,1935+2154 and emit a future FRB. To get a clue, we compare the detected pulses to the radio emission from SGR\,1935+2154.

The energy distributions of GPs from neutron stars and repeating FRBs share a common property, as both are expected to follow a power law. As mentioned in Sect. \ref{sec:ajustes}, we fitted a power law to the single pulse properties, with fluences ranging from $9 \pm 3$\fluence to $58 \pm 10$\fluence. If, as discussed earlier, the intrinsic widths of the pulses are broader, the fluence values would increase. The measured fluences overlap with the faintest values reported in \cite{kirsten_detection_2021} for two bright radio bursts from SGR\,1935+2154, neither those bursts nor the detected GPs in this work show distinct features in their dynamic spectra. The bursts reported by \cite{kirsten_detection_2021} differ from our brightest pulses in two orders of magnitude in their specific luminosities, their waiting time was a fraction of the rotational period and the bursts appeared at random phases, distinguishing them even from FRB20200428, which occurred in opposition to the pulsar-like emission \citep{zhu_radio_2023}. In contrast, we detected pulses at different rotations but within a limited phase range, marking their GP nature. 

Our fluence values are comparable to two bursts reported on 8 October 2020 in \cite{giri_comprehensive_2023}, $(34.1\pm 8.1)$\fluence, $(5.9\pm1.7)$\fluence, which exhibit features in their dynamic spectra and show no phase preference \citep{zhu_radio_2023}. The pulsar-like emission reported by \cite{zhu_radio_2023} shows no evidence of a GP emission at those observing epochs. Their average spectral luminosity is $(6.2\pm0.2) \times 10^{29}\mathrm{\, erg \, s^{-1}}$,  being $10^{-5}$ times the spin down luminosity, in contrast to our value of $10^{-2}$ times. The absence of GP emission but the presence of FRB-like bursts from SGR\,1935+2154 marks a difference in the phenomenology observed from this source with \xte. \xte has shown to be a proficient GP emitter; in contrast, SGR\,1935+2154 has shown high-energy pulses only in the form of FRB and FRB-like. It remains uncertain if SGR\,1935+2154 could emit GPs similar to \xte, or if GPs from \xte could extend to higher luminosities.

In Fig.\,\ref{fig:trans_spacs} we compare the reported single pulses with those of previous works of \xte and SGR\,J1935+2154. We adapted the transient phase space plot from \cite{nimmo_burst_2022}, included the reported pulses from \cite{zhu_radio_2023} assuming a distance of $9.5\pm2.9\mathrm{\, kpc}$, and adopted a distance of $2.5\pm0.5\mathrm{\, kpc}$ for \xte \citep{ding_magnetar_2020}. Several points arise from this comparison: (1) our brightest single pulses overlap in specific luminosity$/4\pi$ with the faintest radio bursts reported by \cite{giri_comprehensive_2023}, (2) we detected some of the most luminous pulses from \xte, consistent with the bright pulses reported by \cite{2024A&A...686A.144B}, (3) SGR\,1935+2154 exhibits standard radio single pulse emission and bright bursts, but almost no intermediate pulses, whereas \xte single pulses spans through a broad range of specific luminosities, (4) all the detected pulses from \xte to date lie below the diagonal line corresponding to a constant brightness temperature of $10^{27} \mathrm{\, K}$.

Studies suggest a twist and un-twist effect on the magnetosphere following the outburst. \cite{borghese_x-ray_2021} reported that the delayed increase in the spin-down was consistent with a steady growth of a small twist. Later, \cite{desvignes_freely_2024} argued that the observed decrease in both spin-down and width of single radio pulses pointed to a gradually untwisting of the magnetic field. Combining the geometrical interpretations from both studies, it appears that our line of sight is close to the region on the neutron star's surface responsible for the X-ray emission, which is also the hemisphere from which the radio pulsar emission seems to originate. This interpretation is supported by the small lag between the X-ray and radio peak reported in \cite{borghese_x-ray_2021}. This scenario is different from the geometry proposed in \cite{zhu_radio_2023} for SGR\,J1935+2154, where the radio pulsar emission and the X-ray peak were in opposition, while FRB 20200428 appeared coincidental with the X-ray peak.

Our study presents that the radio emission from magnetar \xte shows similarities to some intermediate bursts from SGR\,1935+2154, but also important differences, making a direct extrapolation of the possibility of \xte to emit a FRB-like burst difficult. While reviewing this manuscript, a recent work making an in-depth analysis of this possibility has also appeared \citep{lal_low-energy_2025}\footnote{We emphasize the complementary role of observing from the IAR. While that recent work reports observations with the Green Bank Telescope (GBT) (which shares a common visibility window of this magnetar with the IAR), none are reported between MJD 59896 and 60058 (162 days). During this period, we conducted 84 observations, detecting GPs in 35. The peak activity (MJD 59969) occurred during this GBT observational gap, reinforcing the importance of high-cadence IAR observations, as the events reported here are unique and would otherwise have gone undetected.}.

\section{Summary and conclusions}

We have presented the first analysis of \xte observations made with the radio telescopes of the IAR, opening a new line of research for the observatory. We performed a high-cadence campaign during a time window not previously explored, starting four years after the magnetar's latest outburst in 2018. We characterized the emission of GPs during this period and investigated possible X-ray activity.

We detected a total of 249 GPs over 149 days of observations. The 249 GPs were concentrated in 56 out of the 149 observations. We found pulses with flux densities from $16\pm3$~Jy up to $119\pm19$~Jy, centered at a DM=$178.8\pm0.1$\dm. The broad flux density range and the daily fluctuations in the rate of detected pulses highlight the extreme variability in the radio emission from the magnetar.

The GPs we detected share similar specific luminosities with some intermediate bursts from SGR\,1935+2154 reported by \cite{giri_comprehensive_2023}. However, key differences exist with FRB-like pulses from SGR\,1935+2154: the bursts reported by \cite{giri_comprehensive_2023} are not restricted in phase, whereas our detections show a GP phase alignment, confined to a narrow rotational window. This contrasts with FRB and FRB-like emission detected from SGR\,1935+2154, which either appeared in opposition to the pulsar-like component or at random phases \citep{zhu_radio_2023}.

We found that the rate of pulses peaked on 20 January 2023, followed by a sudden drop in the emission of GPs, suggesting a transition from a high-activity state to a lower one. This behavior is coincident with the one reported in \cite{\bause}, where a sudden change in the activity of the magnetar was detected in September 2020 and another one in February 2021. This may imply transitions of the magnetar between different emission states, undergoing a highly active state for two years between February 2021 and February 2023, at which the magnetar's emission fell below our sensitivity.

Finally, we analyzed public archival X-ray observations of the magnetar from \textit{MAXI}. We did not detect any X-ray activity throughout the entire observation period.

\begin{acknowledgements}
     We thank the Referee for their constructive comments, which have significantly improved the quality of this work. We are grateful to the authorities and technical staff of the IAR for their continuous efforts that enable our high-cadence observational campaign. This research has made use of the \textit{MAXI} data provided by RIKEN, JAXA and the \textit{MAXI} team. S.B.A.F. especially thanks M. Tatehiro for the support provided when analysing \textit{MAXI}'s data. S.B.A.F. also thanks M. Lower for providing an initial ephemeris for XTE J1810-197, and M. Bause for the enriching email exchange. S.B.A.F. and E.Z. are PhD candidates with CONICET fellowships. G.E.R. and F.G. acknowledge financial support from the State Agency for Research of the Spanish Ministry of Science and Innovation under grant PID2022-136828NB-C41/AEI/10.13039/501100011033/, and by
     “ERDF A way of making Europe”, by the “European Union”. G.E.R. also thanks the support from PIP 0554 (CONICET). G.E.R and F.G. are CONICET researcher. F.G. acknowledges support from PIBAA 1275 and PIP 0113 (CONICET). S.d.P. acknowledges support from ERC Advanced Grant 789410. COL gratefully acknowledges support from NSF awards AST-2319326, PHY-2207920 and PHY-2513442.
\end{acknowledgements}


\bibliographystyle{aa} % style aa.bst
\bibliography{references,biblio}  % your references biblio.bib

\end{document}